\documentstyle[prd,aps,floats]{revtex}

\def\be{\begin{equation}}
\def\ee{\end{equation}}
\def\ben{\begin{eqnarray}}
\def\een{\end{eqnarray}}

\begin{document}


\input epsf
\renewcommand{\topfraction}{1.0}
\twocolumn[\hsize\textwidth\columnwidth\hsize\csname 
@twocolumnfalse\endcsname

\title{Magnetic dipole moment of the electron}

\author{Franz E.~Schunck\footnote{E-mail: fs@thp.uni-koeln.de}}

\address{Institut f\"ur Theoretische Physik, Universit\"at zu
K\"oln, 50923 K\"oln, Germany}

\date{\today}

\maketitle

\begin{abstract}
We present a model which determines the correct value of the magnetic
dipole moment of the electron. By this, we find a physical meaning for
the electron spin.
\end{abstract}





\vskip2pc]


The attempt to understand the internal structure of the electron has a long
tradition and goes back to Hicks pure geometrical idea of vortices \cite{hicks}
but without explicit relation to the electron.
A few years later, already in 1907, the electron as ringlike axial structure
was introduced, i.e., electromagnetic energy circles around an axis
\cite{stark}. In \cite{lorentz}, the idea of a deforming electron is given
the magnetic moment of which stays nevertheless constant.
Using the experimental results of Stern and Gerlach \cite{stern}, Uhlenbeck and
Goudsmith \cite{uhlen} introduced the idea of an intrinsic angular momentum
of the electron which is twice as large as one would expect classically
and results in the Land\`e or g-factor. Dirac \cite{dirac} followed from the
linearised relativistic Schr\"odinger equation for an electron in a magnetic
field that both an anti-particle (the positron) should exist and
the exact value of the magnetic moment of the electron.

There are a lot of theories which try to understand the internal structure
of elementary particles in general. There exist the ideas of screwvortices
(the {\em archon}) \cite{wiener}, left and right archons together
form a particle.
Broglie applied this as {\em m\'ethode de fusion} \cite{broglie}.
However, Born and Peng introduced a basic element called
the {\em apeiron} \cite{born}.
H\"onl \cite{hoenl} introduced an internal motion of the
electron and Schr\"odinger \cite{schroedinger} a trembling motion.
H\"onl stressed that an electron could be a {\em pole-dipole} particle.
Jehle considered elementary loops which in superposition form electron or quark
\cite{jehle}. Pekeris used a hydromechanical model with stationary circulations
\cite{pekeris}. Dahl had the idea of a rotor model with two elements
\cite{dahl} whereas Hughston applied two twistor elements \cite{hughston}.

Following Mack and Petkova \cite{mack}, quarks can be described by
condensed vortices. 
Harari \cite{harari} introduced as basic element the {\em rishon}
so that elementary particles consist of three of them. 
Pavsic et al.~\cite{pavsic} uses Clifford's $4\times 4$ matrix and interprete
their result as pointlike charge the orbit of which is a cylindrical
helical line.
Following Hautot \cite{hautot}, charge and magnetic moment are two objects
of the eigenstructure, spherically for matter and charge,
axially for angular and magnetic moment.
Wasserman \cite{wasserman} presents as electron model a M\"obius band
asymmetrically cut so that two stripes become interlinked, a M\"obius band
becomes interlinked with a triple twisted M\"obius band. In his opinion,
the twisting is correlated with the spin, and the interlinking with mass
and charge.


The relation between an orbital angular momentum ${\bf L}$ and the
magnetic moment ${\bf \mu }$ is given by the gyromagnetic relation $\gamma_L$
\be
{\bf \mu_L} = \gamma_L {\bf L} \; .
\ee
Within the hydrogen atom the $z$-component of the orbital angular
momentum is a multiple of Planck's
reduced constant $\hbar$ and, then, the magnetic moment is equal to Bohr's
magneton $\mu_B=e\hbar/(2 m_e)$; cf.~below.
Hence, we find that the gyromagnetic relation
for an electron ground state orbit is $\gamma_L= e/(2m_e)$.

An electron possesses besides the orbital also a spin angular momentum ${\bf s}$
the $z$-component of which is given by $\pm \hbar/2$ so that we have
\be
{\bf \mu_s} = \gamma_s {\bf s} = g_s \gamma_L {\bf s} \; ,
\ee
where $g_s$ is the gyromagnetic factor. From experiments, it is found
that the intrinsic magnetic moment of the electron has the absolute
value of Bohr's magneton $\mu_B$ as well.
Because the $z$-component of the spin angular momentum ${\bf s}$
is $\hbar/2$, the gyromagnetic relation is now twice the one of
the orbital angular momentum, i.e., $\gamma_s=e/m_e$ (here the absolute
value); cf.~below. This is also stated that
the gyromagnetic factor $g_s=2$. The factor 2 is confirmed by
Dirac's relativistic
description of the electron in a magnetic field but, in the non-relativistic
limit, the electron still keeps an intrinsic angular momentum of $\hbar/2$
\cite{bohm}.
So far, it is not understood how the
internal charge distribution of the electron could be so that this
result can be explained.

There might be a straightforward solution. Let us remind us that, for an
orbital angular momentum, the total charge is measured from outside.
However, for a spin angular momentum, it becomes important what kind of
{\em internal} charge distribution is present. Before we continue,
let us consider the following.

Bohr's magneton consists of three elementary components ($e, \hbar, m_e$)
which cannot be changed. We can assume that $\hbar$ and the total
electron mass is fixed because the first is 
a universal constant and the latter is measured from outside (but 
can be distributed in some way inside the electron).
But already the electric charge $e$ inside a quark shows us that one
third or two third of the electron charge exist. We conclude that the
electric charge can be arranged in some way.

{}From the measurement of a {\em magnetic dipole moment} we can derive two
characteristics. {\em Dipole} means
that there should exist two opposite electric charges inside the electron.
In order to measure the electric charge $-1e$ outside, we can divide
this up into two electric charges with
$-2e$ and $+1e$. {\em Magnetic} means that at least one charge is moving.

Now we calculate the magneton semi-classically for this
quantenmechanical model. We assume that the charge 
$-2e$ is circling around a central charge $+1e$ moving with radius $r$
and around an area $A$, with velocity $v$, producing a current $I$
\be
\mu = I A = \frac{2ev}{2\pi r} \pi r^2 = e v r
    = \frac{e|{\bf s}|}{m_e} = \frac{e \hbar}{2 m_e} = \mu_B
\ee
where the intrinsic angular momentum $\bf s$ of the double charge, i.e.~the spin
angular momentum, has the quantum mechanical value $\hbar/2$.

We can now rewrite the gyromagnetic relation of the intrinsic magnetic
momentum of the electron by $\gamma_s = e/m_e = 2e/(2m_e)$. Then, we
recognize that the double charge of our model just produces the gyromagnetic
relation as it is observed.
But, we can give the g-factor a physical meaning.
Now $g_s$ is the absolute value of the moving electric charge surrounding
the central electric charge, i.e., $g_s=|-2|$.
Of course, in our simple model, we cannot clarify the exact value of the
gyromagnetic factor as found by quantum electrodynamics.

So far, we ignored the sign of the electric charge. Because the double
charge is negative, we can extract it and find $\mu_s=-\gamma_s {\bf s}$
exactly what is found, namely that the spin angular momentum is anti-parallel
to the intrinsic magnetic momentum.

Clearly, the spin of the electron can be described by our model
simply by visualizing that the double electric charge $-2e$ surrounds
the single central electric charge $+1e$ in an area perpendicular to
the direction of motion either clockwise or anti-clockwise
suggesting spin-up or spin-down, respectively.

We conclude: Our model shows that inside the electron there could be 
a system of two opposite electric charges, desirable would be informations on 
further characteristics of these particles, e.g., whether they are
bosonic or fermionic. But we should be careful. Already the
imagination of a particle (as an electron or a quark) may be wrong.
It might be that the double
charge and the single charge are distributed in some way that the
meaning ``particle'' is no longer tenable. But first let us think of
two particles. Then, we could describe the system quantum mechanically
by using the well-known
hydrogen atom results (with a double charged ``electron'') if we assume 
spherical symmetry. Or due to the idea of a double charge moving
perpendicular to the direction of motion, a cylindrically symmetric system makes
more sense \cite{fluegge}. In both cases, we receive an infinite
amount of energy levels. But, additionally, we know that an
electron possesses no excited states. From this, we conclude that
the two charges inside the electron
are not able to interact via photons.
So, we postulate that we have here a new kind of particle which
carries electric charge but does not interact electromagnetically.
Only the composition of both particles act (as electron) electromagnetically.
Because it is known that even a charged scalar boson can interact
electromagnetically \cite{schunck}, we have here a new kind of
particle which is neither bosonic nor fermionic.
Furthermore, the question arises whether a quantum mechanical description
for this system is allowed even if both particles carry an electric charge.
Alternatively, we can doubt that we have here particles at all but rather
some charge distribution which can be interpreted as particle.

The quantitative mass distribution inside the electron is clear
following our model. We can assume that the electron mass
is mainly concentrated in the central single electric charge and that
the mass of the circulating double electric
charge could be, for example, one thousandth of the central electric
charge as in the hydrogen atom, but only experiments will be able to
determine this.


In this letter, we have introduced a new structure of the electron:
two opposite electric charges inside the electron; we assume that these objects
possess electric charge but do not interact electromagnetically
via photons otherwise we would find a fine-structure inside the electron.
The model explains the magnetic dipole moment and, hence, give us
some hint what spin may be.

\section*{Acknowledgments}

We would like to thank Friedrich W.~Hehl and Eckehard W.~Mielke for discussions.

\end{document}